\documentclass[pre,aps,preprint,showpacs,preprintnumbers,amsmath,amssymb,secnumroman,eqsecnum,eqappnum]{revtex4}

\usepackage{graphicx}% Include figure files
\usepackage{dcolumn}% Align table columns on decimal point
\usepackage{bm}% bold math

\newcommand{\beq}{\begin{equation}}
\newcommand{\eeq}{\end{equation}}
\newcommand{\beqa}{\begin{eqnarray}}
\newcommand{\eeqa}{\end{eqnarray}}
\newcommand{\vc}[1]{\mbox{\boldmath $#1$}}

\newcommand{\vol}[1]{{\bf #1}}

\newcommand{\du}[1]{{\bf\sf #1}}

\begin{document}
%\preprint{APS/123-QED}

\title{Collinear velocity relaxation of two spheres in a viscous incompressible fluid}% Force line breaks with \\

\author{B. U. Felderhof}

 \email{ufelder@physik.rwth-aachen.de}
\affiliation{Institut f\"ur Theorie der Statistischen Physik\\ RWTH Aachen University\\
Templergraben 55\\52056 Aachen\\ Germany\\
}%

%\author{R. B. Jones}

 %\email{r.b.jones@qmul.ac.uk}
%\affiliation{Queen Mary University of London, The School of
%Physics and Astronomy, Mile End Road, London E1 4NS, UK\\}%

\date{\today}% It is always \today, today,

\begin{abstract}
Collinear velocity relaxation of two spheres immersed in a viscous incompressible fluid is studied on the basis of an approximate expression for the retarded hydrodynamic interaction. After a sudden impulse applied to one sphere, the other one instantaneously starts to move as well, with amplitude determined by the added mass effect. The velocities of both spheres eventually decay with a $t^{-3/2}$ long-time tail, but the relative velocity decays with a $t^{-5/2}$ long-time tail. The three relaxation functions are approximated by simple expressions involving only a small number of poles in the complex square root of frequency plane.
\end{abstract}

\maketitle
\section{\label{I}Introduction}
In recent work \cite{1} we investigated the retarded hydrodynamic interaction between two spheres immersed in a viscous incompressible fluid. The corresponding frequency-dependent mobility matrix relates the translational and rotational velocities of the spheres to the hydrodynamic forces and torques exerted by the fluid. In combination with Newton's equations the mobility matrix yields the sphere velocities for oscillatory forces and torques applied to the spheres.

By Fourier analysis the mobility matrix also provides information on the behavior in time of the velocity relaxation functions after a sudden impulse or twist applied to one of the spheres. Due to the assumption of incompressibility there is an instantaneous transfer of momentum, and the other sphere starts to move as well. The nature of the motion provides information on the nature of the fluid. The two-sphere system can be used as an investigative tool to analyze the viscoelasticity of the fluid \cite{2}-\cite{5}.

In the following we consider in particular the translational velocities for collinear motions of the two spheres. For motions transverse to the line of centers there is translation-rotation coupling, leading to a complicated overall motion. The restriction to collinear motion allows a relatively simple one-dimensional picture, but even for this case the motion is intricate.

 The initial values of the sphere velocities after a sudden impulse from a state of rest are determined by added mass effects. We compare the initial values found from the high-frequency behavior of the mobility matrix with those predicted by potential flow theory \cite{6}. The velocity autocorrelation functions, which describe the time-dependent velocities after the initial impulse, decay with a wide distribution of relaxation times. The low-frequency dependence of the mobility matrix incorporates the long-time decay. It is known that the amplitude of the long-time $t^{-3/2}$ decay of the two sphere velocities is identical to that of a single sphere \cite{7}. We show that the relative velocity decays with a $t^{-5/2}$ long-time tail.

Our analysis is based on the recently derived approximate expression for the frequency-dependent mobility matrix \cite{1}. In the approximation the hydrodynamic interaction between the two spheres is limited to a single Green function, but finite size effects are fully taken into account via the exact expression for the primary frequency-dependent Stokes flow, and a Fax\'en theorem in the calculation of the secondary velocity. The approximation should be accurate, unless the spheres initially are very close.

The mobility function for collinear motion is a complicated function of frequency. We show for a numerical example that it can be well approximated by a much simpler function, involving only a small number of poles in the complex square root of frequency plane. In this manner the relaxation behavior fits a general framework of slow dynamics of linear relaxation systems \cite{8}.

The dynamics of two colloidal spheres immersed in a compressible fluid was studied in computer simulation by Tatsumi and Yamamoto \cite{9}. In their theoretical analysis these authors used a simple Green function approximation to the mobility function \cite{10}. In most practical applications it will be sufficient to consider the incompressible limit. Nonetheless it would be of interest to extend the present theory to a compressible fluid.

\section{\label{II}Pair of interacting spheres}

We consider two uniform spheres, labeled $A$ and $B$, with radii $a$ and $b$ and mass densities $\rho_A,\;\rho_B$, immersed in a viscous incompressible fluid, and oscillating with frequency $\omega$ about positions $\vc{R}_A$ and $\vc{R}_B$, as shown in Fig. 1. We choose a Cartesian system of coordinates such that the $z$ axis is along the vector $\vc{R}=\vc{R}_B-\vc{R}_A$. We may choose the origin at $\vc{R}_A$. The fluid has mass density $\rho$ and shear viscosity $\eta$. The fluid flow velocity $\vc{v}(\vc{r},t)$ and pressure $p(\vc{r},t)$ satisfy the linearized Navier-Stokes equations,
 \begin{equation}
\label{2.1}\rho\frac{\partial\vc{v}}{\partial t}=\eta\nabla^2\vc{v}-\nabla p,\qquad\nabla\cdot\vc{v}=0.
\end{equation}
We can linearize, since the oscillations are assumed to be small, so that the inertial Reynolds term can be omitted on the left hand side of the first equation. At the surface of the spheres the fluid velocity and pressure are assumed to satisfy mixed-slip boundary conditions with slip coefficients $\xi_A,\;\xi_B$, respectively \cite{11}. The value $\xi=0$ corresponds to no-slip, and $\xi=1/3$ corresponds to perfect slip \cite{11A}. The fluid fills all space outside the spheres, and is at rest at infinity.

The whole system is caused to move by oscillatory applied forces $\vc{E}_A,\;\vc{E}_B$ and torques
$\vc{N}_A,\;\vc{N}_B$, acting on the spheres, which force the system to oscillate at frequency $\omega$. With the above assumptions the problem is linear. As a consequence the translational velocity amplitudes $\vc{U}_A,\;\vc{U}_B$ and the rotational velocity amplitudes $\vc{\Omega}_A,\;\vc{\Omega}_B$ are linear in the amplitudes of the applied forces and torques. They are also linear in the hydrodynamic forces $\vc{K}_A,\;\vc{K}_B$ and torques $\vc{T}_A,\;\vc{T}_B$ exerted by the fluid on the spheres. The equations of motion for the spheres read in complex shorthand notation
\begin{eqnarray}
\label{2.2}-i\omega\du{m}_p\cdot\du{U}&=&\du{E}+\du{K},\nonumber\\
-i\omega\du{I}_p\cdot\du{\Omega}&=&\du{N}+\du{T},
\end{eqnarray}
with six-vectors $\du{V}=(\vc{V}_A,\vc{V}_B)$. The mass matrix $\du{m}_p$ is a diagonal $6\times 6$ matrix incorporating the two masses $m_A,\;m_B$, and similarly $\du{I}_p$ is a diagonal $6\times 6$ matrix incorporating the two moments of inertia $I_A,\;I_B$. The $12\times 12$ mobility matrix $\vc{\mu}$ is defined by the linear relation
\begin{eqnarray}
\label{2.3}\du{U}&=&-\vc{\mu}^{tt}\cdot\du{K}-\vc{\mu}^{tr}\cdot\du{T},\nonumber\\
\du{\Omega}&=&-\vc{\mu}^{rt}\cdot\du{K}-\vc{\mu}^{rr}\cdot\du{T}.
\end{eqnarray}
The $12\times 12$ friction matrix is the inverse $\vc{\zeta}=\vc{\mu}^{-1}$. The $tt$ part of this matrix has the property that it varies in proportion to $\omega$ at high frequency. The property is called acceleration reaction by Batchelor \cite{12}, who discusses it for a single rigid body. The $6\times 6$ added mass matrix $\du{m}_a$ is defined by
 \begin{equation}
\label{2.4}\vc{\zeta}^{tt}=-i\omega\du{m}_a+{\vc{\zeta}^{tt}}',
\end{equation}
where ${\vc{\zeta}^{tt}}'$ is the remaining part, which does not increase in proportion to $\omega$ at high frequency. It follows from the reciprocal theorem that the various matrices are symmetric \cite{12}. The elements of the mobility matrix define scalar mobility functions. In particular the $tt$ part of the matrix takes the form
\begin{equation}
\label{2.5}\vc{\mu}^{tt}=\left(\begin{array}{cc}\vc{\mu}^{tt}_{AA}&\vc{\mu}^{tt}_{AB}
\\\vc{\mu}^{tt}_{BA}&\vc{\mu}^{tt}_{BB}
\end{array}\right),
\end{equation}
with self-mobility tensors $\vc{\mu}^{tt}_{AA},\vc{\mu}^{tt}_{BB}$ and mutual mobility tensors $\vc{\mu}^{tt}_{AB},\vc{\mu}^{tt}_{BA}$ given by
\begin{equation}
\label{2.6}\vc{\mu}^{tt}_{ij}=\alpha^{tt}_{ij}(R,\omega)\hat{\vc{R}}\hat{\vc{R}}+\beta^{tt}_{ij}(R,\omega)(\vc{I}-\hat{\vc{R}}\hat{\vc{R}}),
\end{equation}
with $\hat{\vc{R}}=\vc{R}/R$, and scalar mobility functions $\alpha^{tt}_{ij}(R,\omega)$ and $\beta^{tt}_{ij}(R,\omega)$. The form follows from translational and rotational invariance. Translation-rotation coupling is expressed by scalar mobility functions $\beta^{tr}_{ij}(R,\omega)$ and $\beta^{rt}_{ij}(R,\omega)$, which are related by the reciprocity relation $\beta^{tr}_{ij}(R,\omega)=-\beta^{rt}_{ji}(R,\omega)$. The tensor $\vc{\mu}^{tr}_{ij}$ takes the form
\begin{equation}
\label{2.7}\vc{\mu}^{tr}_{ij}=\beta^{tr}_{ij}(R,\omega)\vc{\epsilon}\cdot\hat{\vc{R}},
\end{equation}
where $\vc{\epsilon}$ is the Levi-Civita tensor. The tensor $\vc{\mu}^{rr}_{ij}$ takes a form similar to Eq. (2.6).

We consider first the longitudinal case, where both spheres translate along the $z$ axis, parallel to $\hat{\vc{R}}$. Then by symmetry there is no translation-rotation coupling, and it suffices to consider the scalar mobility functions $\alpha^{tt}_{ij}(R,\omega)$. Two of these are related by the reciprocity relation $\alpha^{tt}_{AB}(R,\omega)=\alpha^{tt}_{BA}(R,\omega)$, and $\alpha^{tt}_{BB}(R,\omega)$ can be obtained from $\alpha^{tt}_{AA}(R,\omega)$ by $AB$ interchange.

Elsewhere we derived an approximate expression for the mutual mobility function $\alpha^{tt}_{AB}(R,\omega)$ based on a one-propagator approximation, which takes account of only a single Green function between the two spheres \cite{1}. The primary flow is a Stokes flow at frequency $\omega$ generated by sphere $A$ as if it were moving by itself in infinite fluid. The velocity of sphere $B$ as it moves in this flow with zero force is calculated from a Fax\'en theorem \cite{13}. The resulting mutual mobility function reads \cite{1}
\begin{equation}
\label{2.8}\alpha^{tt}_{BA}(R,\omega)=\frac{B_0(\alpha a,\xi_A)B_0(\alpha b,\xi_B)-(1+\alpha R)e^{\alpha (a+b-R)}}{2\pi\eta\alpha^2R^3A_0(\alpha a,\xi_A)A_0(\alpha b,\xi_B)},
\end{equation}
with the abbreviations
\begin{equation}
\label{2.9}A_0(\lambda,\xi)=(1-\xi)\frac{1+\lambda}{1+\xi\lambda}+\frac{1}{9}\lambda^2,\qquad B_0(\lambda,\xi)=(1-\xi)\frac{1+\lambda}{1+\xi\lambda}+\frac{1}{3}\lambda^2,
\end{equation}
and $\alpha=\sqrt{-i\omega\rho/\eta}$, $\mathrm{Re}\;\alpha>0$.
The approximate self-mobility function $\alpha^{tt}_{AA}(R,\omega)$ is more complicated \cite{1}. It is calculated from a single reflection from sphere $B$, which is freely moving. The effect of the correction to the single particle mobility $\mu^t_A=1/\zeta^t_A$ on the velocity relaxation function is numerically small for not too close distances.
In our analysis we use the approximation $\alpha^{tt}_{AA}(R,\omega)\approx 1/\zeta^t_A(\omega,\xi_A)$ with single sphere friction coefficient
\begin{equation}
\label{2.10}\zeta^t_A(\omega,\xi_A)=6\pi\eta aA_0(\alpha a,\xi_A).
\end{equation}
We can estimate the validity of the approximation by considering the added mass matrix which follows from these expressions. We find that the part corresponding to longitudinal motions is given by
\begin{equation}
\label{2.11}\du{m}^\parallel_a=\frac{2\pi\rho}{3}\bigg[1+O(\frac{a^3b^3}{R^6})\bigg]\left(\begin{array}{cc}a^3&-3\frac{a^3b^3}{R^3}
\\-3\frac{a^3b^3}{R^3}&b^3
\end{array}\right).
\end{equation}
The terms omitted in $\alpha^{tt}_{AA}(R,\omega)$ contribute to the $O(a^3b^3/R^6)$ term. The diagonal terms follow from the added mass of each individual sphere \cite{12},\cite{13}.

In the same one-propagator approximation we found for the mutual mobility function for motions normal to the line of centers
\begin{equation}
\label{2.12}\beta^{tt}_{BA}(R,\omega)=\frac{-(1+\xi_A\alpha a)(1+\xi_B\alpha b)B_0(\alpha a,\xi_A)B_0(\alpha b,\xi_B)+(1-\xi_A)(1-\xi_B)A_1(\alpha R)e^{\alpha(a+b-R)}}{4\pi\eta\alpha^2R^3(1+\xi_A\alpha a)(1+\xi_B\alpha b)A_0(\alpha a,\xi_A)A_0(\alpha b,\xi_B)},
\end{equation}
with function
\begin{equation}
\label{2.13}A_1(\lambda)=1+\lambda+\lambda^2.
\end{equation}
At long range it is not necessary to consider the translation-rotation coupling, since at finite frequency the corresponding mobility functions decay exponentially. In the same way as above we find for the part of the added mass matrix corresponding to transverse motions
\begin{equation}
\label{2.14}\du{m}^\perp_a=\frac{2\pi\rho}{3}\bigg[1+O(\frac{a^3b^3}{R^6})\bigg]\left(\begin{array}{cc}a^3&\frac{3a^3b^3}{2R^3}
\\\frac{3a^3b^3}{2R^3}&b^3
\end{array}\right).
\end{equation}
Note that the added mass matrices in Eqs. (2.11) and (2.14) are independent of the two slip coefficients.

The expressions for the added mass matrices are identical with the ones derived by Lamb \cite{14} in potential flow theory from the kinetic energy of the flow pattern. The results also follow from a linear response theory \cite{6} based on the expression for the force on a sphere subjected to an incident potential flow, as derived by Landau and Lifshitz \cite{15} and by Batchelor \cite{12}. In dipole approximation this leads to an expression for the added mass matrix given by \cite{16}
\begin{equation}
\label{2.15}\du{m}_a=-\du{m}_f+4\pi\rho\vc{\mathcal{A}},
\end{equation}
where the matrix $\du{m}_f$ is diagonal with elements $m_{fA}=4\pi\rho a^3/3,\;m_{fB}=4\pi\rho b^3/3$ corresponding to the displaced mass of each sphere, and $\vc{\mathcal{A}}$ is the inverse of a matrix involving the interactions between induced dipoles. The evaluation of Eq. (2.15) for two spheres \cite{17} yields results consistent with Eqs. (2.11) and (2.14).

\section{\label{III}Velocity relaxation}

Added mass affects influence the response of the system to a sudden impulse applied to one of the spheres. Let the fluid and spheres be at rest for $t<0$, and consider small applied forces of the form $\du{E}(t)=\du{S}\delta(t)$ with impulse vector $\du{S}=(\vc{S}_A,\vc{S}_B)$. For $\vc{S}_B=0$ both sphere $A$ and sphere $B$ start to move at $t=0+$. More generally we have
\begin{equation}
\label{3.1}\du{U}(0+)=\du{m}^{-1}\cdot\du{S},
\end{equation}
with six-dimensional mass matrix $\du{m}=\du{m}_p+\du{m}_a$. The equation generalizes the known acceleration reaction for a single sphere [13] to two spheres. The sudden push on both spheres creates an irrotational flow pattern with boundary values corresponding to the two sphere velocities [14]. The linear relation between sphere velocities and the imposed impulses follows from the pressure exerted on each sphere and defines the mass matrix [6].

At later times
\begin{equation}
\label{3.2}\du{U}(t)=\du{R}(t)\cdot\du{S},\qquad t>0,
\end{equation}
with a relaxation matrix $\du{R}(t)$ which has the one-sided Fourier transform
\begin{equation}
\label{3.3}\hat{\du{R}}(\omega)=\int^\infty_0e^{i\omega t}\du{R}(t)\;dt.
\end{equation}
We can identify
\begin{equation}
\label{3.4}\hat{\du{R}}(\omega)=\du{Y}^{tt}(\omega),\qquad\du{Y}(\omega)=\big[-i\omega\du{M}_p+\vc{\zeta}(\omega)\big]^{-1},
\end{equation}
with $12\times 12$ generalized mass matrix $\du{M}_p$ which follows from Eq. (2.2). The matrix $\du{Y}(\omega)$ is called the admittance matrix \cite{19}.

The situation is simplest for longitudinal motions. If sphere $A$ gets a sudden push in the direction of the line of centers, then also sphere $B$ starts to move in the same direction. The reaction is instantaneous due to the assumption of incompressibility. In a compressible fluid the reaction would take some time due to the finite velocity of sound, as seen in the computer simulation of Tatsumi and Yamamoto \cite{9}. From Eq. (3.4) we can evaluate how the velocity of each sphere relaxes after the initial push. It follows from a general theorem derived by Cichocki and Felderhof \cite{7} that at long times the velocity of each sphere decays with a $t^{-3/2}$ long-time tail, with an amplitude which is the same as if each sphere were by itself. The initial values of the two velocities are determined by the effective mass matrix, including the added mass terms which come from the high frequency behavior of the friction matrix $\vc{\zeta}^{tt}(\omega)$. Neither the initial values, nor the amplitude of the long-time tails, depend on the slip coefficients $(\xi_A,\xi_B)$. Since the friction coefficient of a single sphere decreases from the Stokes value $6\pi\eta a$ for no-slip to $6\pi\eta(1-\xi_A)$ with increasing slip coefficient $\xi_A$, we expect that the mean relaxation time will increase when the slip coefficients $(\xi_A,\xi_B)$ increase.

For two Brownian spheres in thermal equilibrium the velocities are not correlated, and the thermal average $<\vc{U}_A\vc{U}_B>$ vanishes. From Eq. (2.11) we see that after a very short time the longitudinal velocity components are correlated positively, whereas Eq. (2.14) shows that then the transverse components are correlated negatively. At later times the velocity relaxation function $\vc{C}_{AB}(t)=<\vc{U}_A(t)\vc{U}_B(0)>$ is related to the relaxation matrix by the fluctuation-dissipation theorem $\vc{C}_{AB}(t)=k_BT\vc{R}_{AB}(t)$.

It is convenient to take the $z$ axis along the line of centers. Then by symmetry the six-dimensional matrix $\du{R}(t)$ decomposes into a two-dimensional matrix $\du{R}^\parallel(t)$ corresponding to longitudinal motions in the $z$ direction, and two identical two-dimensional matrices $\du{R}^\perp(t)$ corresponding to transverse motions in the $x$ and $y$ directions.

We consider scalar autocorrelation functions of the form
\begin{equation}
\label{3.5}C(t)=(\psi|\du{R}(t)|\psi),
\end{equation}
where $|\psi)$ is a chosen six-dimensional vector selecting a linear combination of translational velocity components. We define the one-sided Fourier transform as
\begin{equation}
\label{3.6}\hat{C}(\omega)=\int^\infty_0e^{i\omega t}C(t)\;dt.
\end{equation}
This is given by
\begin{equation}
\label{3.7}\hat{C}(\omega)=(\psi|\du{Y}^{tt}(\omega)|\psi).
\end{equation}
The initial value of the autocorrelation function is
\begin{equation}
\label{3.8}C(0+)=(\psi|\du{m}^{-1}|\psi).
\end{equation}
We write the autocorrelation function in the form
\begin{equation}
\label{3.9}C(t)=C(0+)\gamma(t/\tau_M),
\end{equation}
with initial value $\gamma(0+)=1$ and mean relaxation time
\begin{equation}
\label{3.10}\tau_M=\frac{1}{C(0+)}\int^\infty_0C(t)\;dt.
\end{equation}
From Eqs. (3.7) and (3.8) we find
\begin{equation}
\label{3.11}\tau_M=\frac{(\psi|\vc{\mu}^{tt}(0)|\psi)}{(\psi|\du{m}^{-1}|\psi)}.
\end{equation}
We define the variable $z=-i\omega\tau_M$ and the function
\begin{equation}
\label{3.12}\Gamma(z)=\frac{\hat{C}(\omega)}{(\psi|\vc{\mu}^{tt}(0)|\psi)}.
\end{equation}
This has the properties
\begin{equation}
\label{3.13}\Gamma(0)=1,\qquad\lim_{z\rightarrow\infty} z\Gamma(z)=1.
\end{equation}
Since the functions defined in Eqs. (2.8) and (2.12) depend on frequency via the variable $\alpha$, the dependence of $\Gamma(z)$ on $z$  is via $y=\sqrt{z}$. The spectral density $p(u)$ is defined by \cite{8}
\begin{equation}
\label{3.14}p(u)=\frac{1}{\pi}\;\mathrm{Im}[\Gamma(y\rightarrow -i\sqrt{u})],
\end{equation}
for positive $u$. The relaxation function $\gamma(\tau)$ in Eq. (3.5) is given by the inverse Stieltjes transform \cite{18}
\begin{equation}
\label{3.15}\gamma(t/\tau_M)=\int^\infty_0p(u)\exp[-ut/\tau_M]\;du.
\end{equation}
The function $\Gamma(z)$ has the Stieltjes representation
\begin{equation}
\label{3.16}\Gamma(z)=\int^\infty_0\frac{p(u)}{u+z}\;du.
\end{equation}
Hence the spectral density has the properties
\begin{equation}
\label{3.17}\int^\infty_0p(u)\;du=1,\qquad\int^\infty_0\frac{p(u)}{u}\;du=1,
\end{equation}
corresponding to Eq. (3.13). The second property in Eqs. (3.13) and (3.17) corresponds to the choice of $\tau_M$ as the timescale.

Since we want to compare the spectral densities for different vectors $\psi$ it is more convenient to write the relaxation function as
\begin{equation}
\label{3.18}\gamma(t/\tau_M)=\int^\infty_0P(s)e^{-st}\;ds,
\end{equation}
with rate distribution
\begin{equation}
\label{3.19}P(s)=\tau_Mp(\tau_Ms),
\end{equation}
where $s$ is the relaxation rate $s=u/\tau_M$.

\section{\label{IV}Collinear motion}

We consider velocity relaxation along the line of centers in some more detail. The scalar mobility function in Eq. (2.8) is a complicated function of frequency and correspondingly the various autocorrelation functions are intricate functions of time. We show that, provided the two spheres are of comparable size, a relatively simple approximate description can be found.

It is known that for a single sphere the velocity relaxation function shows an important long-time tail. In that case the spectral density is of the form \cite{19}
 \begin{equation}
\label{4.1}p_2(u)=\frac{1}{\pi}\frac{\sigma\sqrt{u}}{1+(\sigma^2-2)u+u^2},
\end{equation}
where the parameter $\sigma$ can be found from the mass and the added mass of the sphere. The long-time tail corresponds to the square root singularity at $u=0$.
The velocity relaxation function is a sum of two $w$-functions,
  \begin{equation}
\label{4.2}\gamma(\tau)=\frac{1}{\sqrt{\sigma^2-4}}[y_+w(-iy_+\sqrt{\tau})-y_-w(-iy_-\sqrt{\tau})],
\end{equation}
with
  \begin{equation}
\label{4.3}w(z)=e^{-z^2}\mathrm{erfc}(-i\sqrt{z}),
\end{equation}
and values $y_\pm$ which correspond as $y_\pm=\sqrt{z_\pm}$ to the zeros of the denominator of the Laplace transform of the relaxation function,
   \begin{equation}
\label{4.4}\Gamma_2(z)=\frac{1}{1+\sigma\sqrt{z}+z},
\end{equation}
where $z=-i\omega\tau_M$, with mean relaxation time $\tau_M=m^*/(6\pi\eta a)$ for a sphere of radius $a$ and mass $m_0$ with effective mass $m^*=m_0+m_f/2$ and $m_f=4\pi\rho a^3/3$.
The relaxation function has the long-time behavior
  \begin{equation}
\label{4.5}\gamma(\tau)\approx\frac{\sigma}{2\sqrt{\pi}}\;\tau^{-3/2}\qquad\mathrm{as}\qquad\tau\rightarrow\infty,
\end{equation}
where $\tau=t/\tau_M$ and $\sigma=\sqrt{9m_f/2m^*}$. The corresponding long-time behavior of the velocity autocorrelation function is
  \begin{equation}
\label{4.6}C(t)\approx\frac{1}{12}\sqrt{\rho}(\pi\eta t)^{-3/2}\qquad\mathrm{as}\qquad t\rightarrow\infty.
\end{equation}

For the pair of spheres in collinear motion it suffices to consider the two-dimensional matrix $\du{R}^\parallel(t)$ and corresponding two-vectors $|\psi)$. We consider the three vectors
\begin{equation}
\label{4.7}|\psi)_A=(1,0),\qquad|\psi)_B=(0,1),\qquad |\psi)_d=(1,-1).
\end{equation}
The first corresponds to motion of sphere $A$, the second to motion of sphere $B$, and the third to relative motion. We then have
\begin{eqnarray}
\label{4.8}C^\parallel_{AA}(t)&=&R^\parallel_{AA}(t),\qquad C^\parallel_{BB}(t)=R^\parallel_{BB}(t),\nonumber\\
 C^\parallel_{dd}(t)&=&R^\parallel_{AA}(t)-2R^\parallel_{AB}(t)+R^\parallel_{BB}(t).
\end{eqnarray}
This shows that conversely the three elements $R^\parallel_{ij}(t)$ can be found from the three autocorrelation functions,
\begin{eqnarray}
\label{4.9}R^\parallel_{AA}(t)&=&C^\parallel_{AA}(t),\qquad R^\parallel_{BB}(t)=C^\parallel_{BB}(t),\nonumber\\
R^\parallel_{AB}(t)&=& \frac{1}{2}[C^\parallel_{AA}(t)-C^\parallel_{dd}(t)+C^\parallel_{BB}(t)].
\end{eqnarray}
It follows from a general theorem \cite{7} that the relaxation functions $R^\parallel_{AA}(t)$ and $R^\parallel_{BB}(t)$ have exactly the same long-time behavior as in Eq. (4.6). At long times the two spheres move collectively. We show below that the relative velocity decays with a $t^{-5/2}$ long-time tail.

We express the autocorrelation functions as in Eq. (3.17). This yields the rate distributions $P^\parallel_{AA}(s),P^\parallel_{dd}(s),P^\parallel_{BB}(s)$ as functions of the rate $s$. Hence we find the time-dependent autocorrelation functions $C^\parallel_{AA}(t),C^\parallel_{dd}(t),C^\parallel_{BB}(t)$ by numerical integration.

The initial values of the autocorrelation functions, as given by Eq. (3.8), are
\begin{eqnarray}
\label{4.10}C^\parallel_{AA}(0+)&=&\frac{3}{2\pi a^3}\frac{R^6(\rho+2\rho_B)-18a^3b^3\rho_B}{R^6(\rho+2\rho_A)(\rho+2\rho_B)-36a^3b^3\rho_A\rho_B},\nonumber\\
C^\parallel_{dd}(0+)&=&
\frac{3}{2\pi a^3b^3}\frac{a^3R^6(\rho+2\rho_A)+b^3R^6(\rho+2\rho_B)-6a^3b^3R^3\rho-18a^6b^3\rho_A-18a^3b^6\rho_B}{R^6(\rho+2\rho_A)(\rho+2\rho_B)-36a^3b^3\rho_A\rho_B},\nonumber\\
\end{eqnarray}
with $C^\parallel_{BB}(0+)$ found by $AB$-interchange.

The three mean relaxation times $\tau_{MAA},\tau_{Mdd},\tau_{MBB}$ are given by Eq. (3.11) as
\begin{eqnarray}
\label{4.11}\tau^\parallel_{MAA}&=&\frac{a^2}{9\eta}\frac{R^6(\rho+2\rho_A)-18a^3b^3\rho_A}{R^6-9a^3b^3},\nonumber\\
\tau^\parallel_{Mdd}&=&\frac{a^3b+ab^3-3abR^2+aR^3+bR^3}{6\pi\eta abR^3C_{dd}(0+)},
\end{eqnarray}
with $\tau^\parallel_{MBB}$ found by $AB$-interchange.

We note that the two-dimensional admittance matrix has the low-frequency expansion \cite{7}
\begin{equation}
\label{4.12}\du{Y}^{tt\parallel}(\omega)=\vc{\mu}^{tt\parallel}(0)-\frac{\alpha}{6\pi\eta}\left(\begin{array}{cc}1&1
\\1&1
\end{array}\right)+O(\alpha^2a^2).
\end{equation}
This corresponds to the universal nature of the collective long-time motion, mentioned above.

We show below for a numerical example with two spheres of comparable size that the three relaxation functions are well approximated by a sum of $w$-functions similar to Eq. (4.2) with either two or three terms. The simple approximate description corresponds to a small number of poles of the admittance in the complex $\sqrt{z}$ plane.

\section{\label{V}Numerical example}

As an example we consider two neutrally buoyant spheres of radii $a,\;b=2a$, at center-to-center distance $R=5a$, with mass densities $\rho_A=\rho$ and $\rho_B=\rho$, with no-slip boundary conditions, corresponding to slip coefficients $\xi_A=\xi_B=0$. We consider motion along the line of centers. The explicit expression for the admittance matrix $\du{Y}^{tt\parallel}(\omega)$ is quite complicated, but it is straightforward to obtain numerical results. We compare with the simplified description discussed above.

First we evaluate the initial values of the relaxation functions, as given by Eqs. (4.9) and (4.10). This yields
\begin{equation}
\label{5.1}m_A R^{tt\parallel}_{AA}(0+)=0.6660,\qquad m_A R^{tt\parallel}_{BB}(0+)=0.0832,\qquad m_A R^{tt\parallel}_{AB}(0+)=0.0053,
\end{equation}
where $m_A=4\pi\rho a^3/3$. Note that the first value is close to $m_A/m^*_A=2/3$, where $m^*_A$ is the effective mass of a single sphere of radius $a$. The difference from unity is due to the added mass effect in an incompressible fluid. For a compressible fluid the value would be unity, as explained by Zwanzig and Bixon \cite{20}.  Similarly the second value is close to $m_A/m^*_B=1/12$. The difference from zero of the third value is also due to the instantaneous transfer of momentum in an incompressible fluid.

Next we evaluate the mean relaxation times, as given by Eq. (4.11). This yields
\begin{equation}
\label{5.2}\tau_{MAA}=0.33385\tau_{vA},\qquad\tau_{MBB}=1.3354\tau_{vA}\qquad \tau_{Mdd}=0.2828\tau_{vA},
\end{equation}
with viscous relaxation time $\tau_{vA}=a^2\rho/\eta$. The first value is close to the single sphere value $m^*_A/(6\pi\eta a)=\tau_{vA}/3$, and the second value is close to the single sphere value $m^*_B/(6\pi\eta b)=4\tau_{vA}/3$.

The transforms $\Gamma_{AA}(z),\;\Gamma_{BB}(z)$ and $\Gamma_{dd}(z)$ can be found from Eqs. (3.6), (3.12) and (4.7) with in each case the appropriate value $z=-i\omega\tau_{Mjj}$ with $j=A,B,d$. The corresponding spectral densities $p_{AA}(u),\;p_{BB}(u)$ and $p_{dd}(u)$ are found from Eq. (3.14).

In Fig. 2 we plot the spectral density $p_{AA}(u)$ calculated from Eq. (3.14) with $z=-i\omega\tau_{MAA}$. We compare with the two-pole expression $p_{2AA}(u)$ calculated for sphere $A$ from Eq. (4.1) with parameter
 \begin{equation}
\label{5.3}\sigma_{AA}=\sqrt{\frac{9m_{fA}}{2}\;C^\parallel_{AA}(0+)},
\end{equation}
with $C^\parallel_{AA}(0+)$ given by Eq. (4.10).
In Fig. 3 we plot the spectral density $p_{BB}(u)$ calculated from Eq. (3.14) with $z=-i\omega\tau_{MBB}$. We compare with $p_{2BB}(u)$, calculated in the same manner with parameter $\sigma_{BB}$. In both cases the agreement is nearly perfect. The spectral density covers several decades in the dimensionless variable $u$.

We cannot expect the spectral density $p_{dd}(u)$ to be well approximated by a two-pole expression, since the expansion of the function $\Gamma_{dd}(z)$ in powers of $\sqrt{z}$ does not have a term linear in $\sqrt{z}$, as seen from the second term in Eq. (4.12) and the expression $|\psi_d)=(1,-1)$ from Eq. (4.7). Instead we compare with an expression of the form
 \begin{equation}
\label{5.4}\Gamma_3(z)=\frac{1}{1+z+\frac{Cz}{1+D\sqrt{z}}},
\end{equation}
familiar from the theory of viscoelasticity of suspensions \cite{21},\cite{22}, with $z=-i\omega\tau_{Mdd}$. This is a three-pole expression, characterized by three poles in the complex $y=\sqrt{z}$ plane. The parameters $C,D$ can be found by comparison with the expansion of $\Gamma_{dd}(z)$ in powers of $y=\sqrt{z}$.

The expansion of $\Gamma_{dd}(z)$ in powers of $y$,
 \begin{equation}
\label{5.5}\Gamma_{dd}(z)=1+c_2y^2+c_3y^3+O(y^4).
\end{equation}
yields for the coefficients $C$ and $D$
\begin{equation}
\label{5.6}C=-1-c_2,\qquad D=\frac{-c_3}{1+c_2}.
\end{equation}
We find for the coefficient $c_2$
 \begin{equation}
\label{5.7}c_2=\frac{-ab}{36\eta R^3}\;\frac{n_2}{a^3b+ab^3-3abR^2+(a+b)R^3},
\end{equation}
with numerator
\begin{eqnarray}
\label{5.8}n_2&=&2(a^2+b^2)(a^2+b^2-6R^2)(\rho_Aa^3+\rho_Bb^3)\nonumber\\
&+&\big[(8\rho_A-5\rho)a^4+(8\rho_B-5\rho)b^4+2a^2b^2(4\rho_A+4\rho_B-11\rho)\big]R^3\nonumber\\
&+&18 (\rho_A a^3+\rho_B b^3)R^4+6\big[(7\rho-4\rho_A)a^2+(7\rho-4\rho_B)b^2\big]R^5\nonumber\\
&+&8\big[(\rho_A-4\rho)a+(\rho_B-4\rho)b\big]R^6+27\rho R^7,
\end{eqnarray}
and for the coefficient $c_3$
 \begin{equation}
\label{5.9}c_3=\bigg(\frac{\rho}{\eta\tau_{Mdd}}\bigg)^{3/2}\frac{ab}{45}\;\frac{4a^5+4b^5+10a^2b^3+10a^3b^2-30(a^3+b^3)R^2-9R^5}{a^3b+ab^3-3abR^2+(a+b)R^3}.
\end{equation}
In Fig. 4 we plot the spectral density $p_{dd}(u)$ calculated from Eq. (3.14) with $z=-i\omega\tau_{Mdd}$ and relaxation time $\tau_{Mdd}$ given by Eq. (4.11). We compare with $p_{3dd}(u)$, calculated in the same manner from $\Gamma_3(z)$ with parameters $C$ and $D$. In both cases the agreement is nearly perfect. The spectral density covers several decades in the dimensionless variable $u$. The low frequency behavior of the transform $\Gamma_{dd}(z)$ corresponds to a $t^{-5/2}$ long-time tail of the function $C_{dd}(t)$.

In Fig. 5 we plot the rate distributions $P^\parallel_{AA}(s),P^\parallel_{dd}(s),P^\parallel_{BB}(s)$, multiplied by $C^\parallel_{jj}(0+) m_a/\tau_{vA}$ for $j=(A,d,B)$, as functions of $\log_{10}(s\tau_{vA})$, as given by Eq. (3.19) for the respective mean relaxation times. It can be seen from the behavior of the spectra for small relaxation rates that the relaxation functions $R^\parallel_{AA}(t)$ and $R^\parallel_{BB}(t)$ have the same long-time behavior. In Fig. 6 we plot $\log_{10}[m_AR^\parallel_{AA}(t)]$ and $\log_{10}[m_AR^\parallel_{dd}(t)]$ as functions of $\log_{10}(t/\tau_{vA})$. The first function has a $t^{-3/2}$ long-time tail, and the second one has a $t^{-5/2}$ long-time tail.

\section{\label{VI}Discussion}

In the above we studied velocity relaxation of two spheres immersed in a viscous incompressible fluid. In the simplest configuration the motion of both spheres is along the line of centers. The transfer of momentum after an initial impulse applied to one of the spheres is instantaneous due to incompressibility. The initial values of the two velocities in this situation are expressed by an added mass matrix which shows a long range dependence on the distance between centers. The relaxation at subsequent times is complicated and occurs due to viscous diffusion and interference of flow patterns. The velocities of both spheres decay eventually with the same $t^{-3/2}$ long-time tail.

The explicit calculations of the velocity relaxation functions are performed on the basis of an approximate expression for the retarded scalar mobility function derived elsewhere \cite{1}. We expect that for not too near distances the approximation provides accurate results which may be compared with experiment and computer simulation.

We showed that for collinear motion the relaxation functions can be described by a relatively small number of elementary modes with parameters which can be evaluated from the sphere properties. The relaxation functions are characterized conveniently by rate distributions. We expect that other aspects of velocity relaxation of two spheres in hydrodynamic interaction can be described in similar manner.

\newpage

\newpage

\section*{Figure captions}
\subsection*{Fig. 1}
Snapshot of two spheres oscillating along the line of centers.

\subsection*{Fig. 2}
Plot of the spectral density $p_{AA}(u)$ for two spheres $A, B$ as specified at the beginning of Sec. V (solid curve). We compare with the two-pole approximation $p_{2AA}(u)$ (dashed curve).

\subsection*{Fig. 3}
Plot of the spectral density $p_{BB}(u)$ for two spheres $A, B$ as specified at the beginning of Sec. V (solid curve). We compare with the two-pole approximation $p_{2BB}(u)$ (dashed curve).

\subsection*{Fig. 4}
Plot of the spectral density $p_{dd}(u)$ for two spheres $A, B$ as specified at the beginning of Sec. V (solid curve). We compare with the three-pole approximation $p_{3dd}(u)$ (dashed curve).

\subsection*{Fig. 5}
Plot of the three rate distributions $P^\parallel_{AA}(s)$ (solid curve),  $P^\parallel_{dd}(s)$ (short dashes), and $P^\parallel_{BB}(s)$ (long dashes), multiplied by $C^\parallel_{jj}(0+) m_a/\tau_{vA}$ for $j=(A,d,B)$, as functions of $\log_{10}(s\tau_{vA})$.

\subsection*{Fig. 6}
Plot of the functions $\log_{10}[m_AR^\parallel_{AA}(t)]$ (solid curve) and $\log_{10}[m_AR^\parallel_{dd}(t)]$ (dashed curve) as functions of $\log_{10}(t/\tau_{vA})$.

\newpage
\clearpage
\newpage
\setlength{\unitlength}{1cm}
\begin{figure}
 \includegraphics{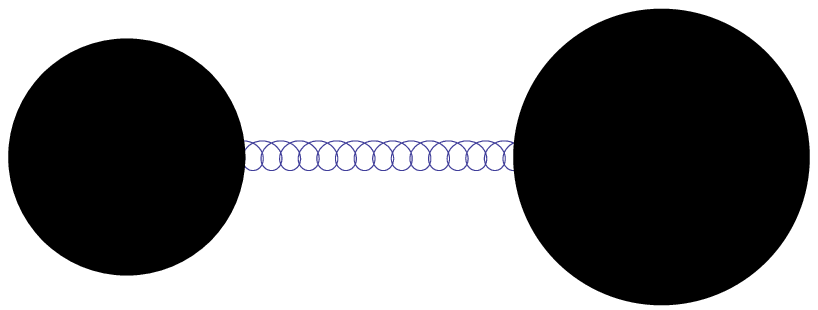}
   \put(-9.1,3.1){}
\put(-1.2,-.2){}
  \caption{}
\end{figure}
\newpage
\clearpage
\newpage
\setlength{\unitlength}{1cm}
\begin{figure}
 \includegraphics{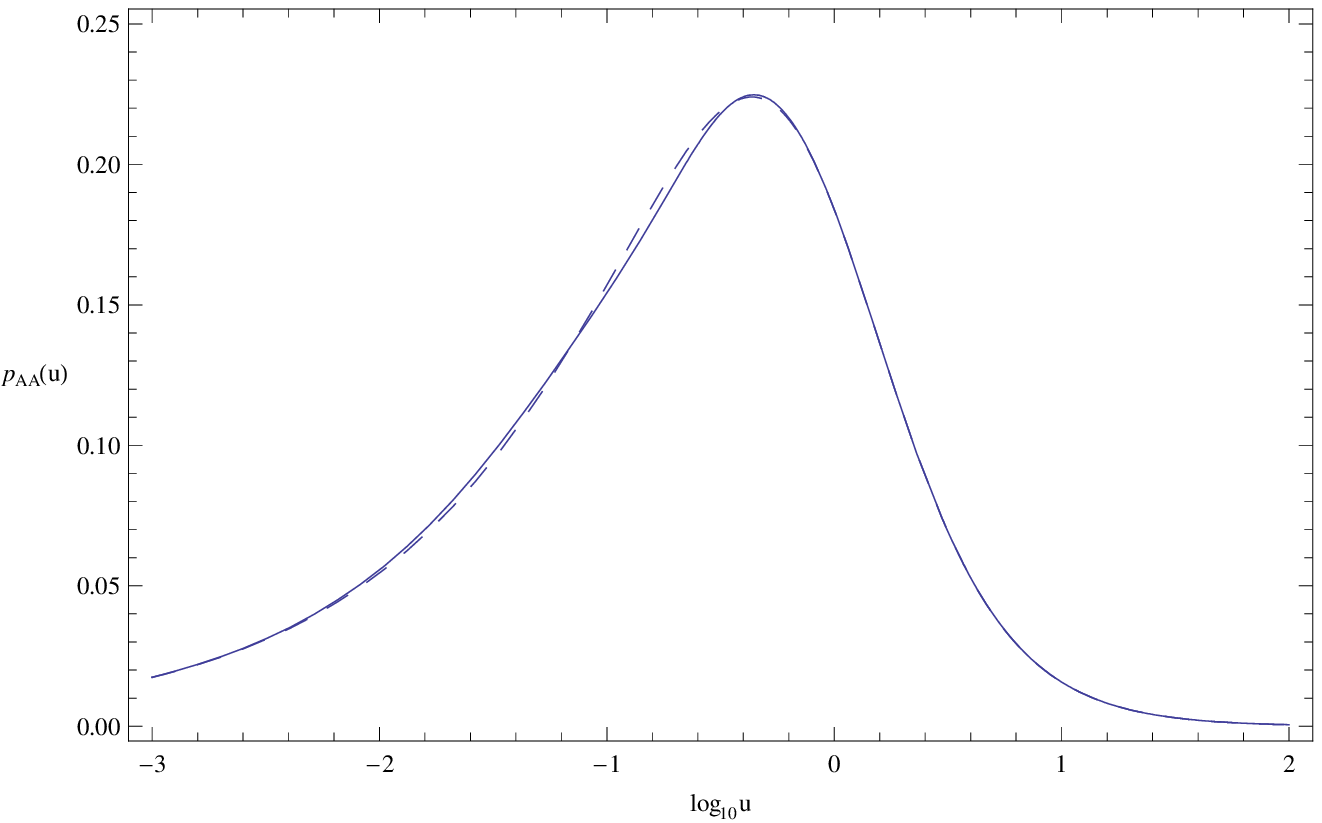}
   \put(-9.1,3.1){}
\put(-1.2,-.2){}
  \caption{}
\end{figure}
\newpage
\clearpage
\newpage
\setlength{\unitlength}{1cm}
\begin{figure}
 \includegraphics{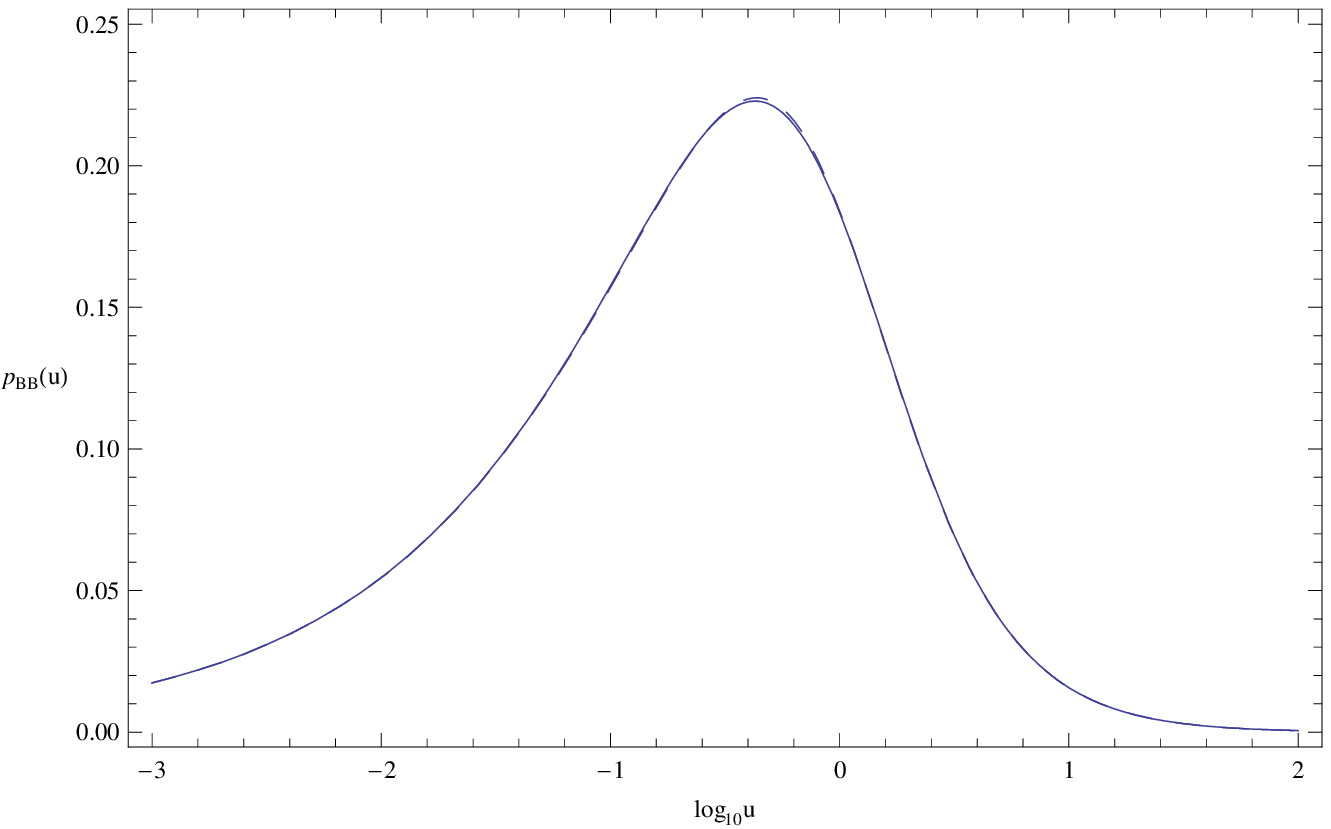}
   \put(-9.1,3.1){}
\put(-1.2,-.2){}
  \caption{}
\end{figure}
\newpage
\clearpage
\newpage
\setlength{\unitlength}{1cm}
\begin{figure}
 \includegraphics{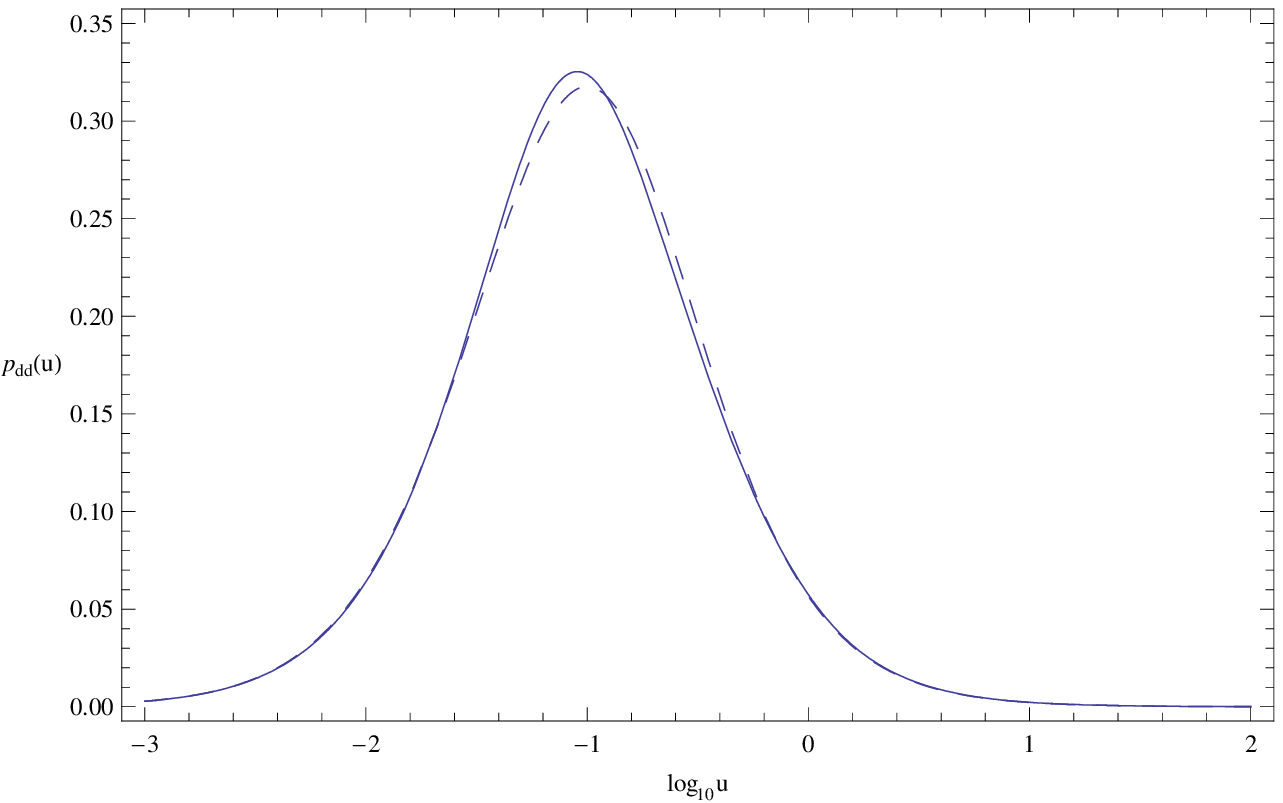}
   \put(-9.1,3.1){}
\put(-1.2,-.2){}
  \caption{}
\end{figure}
\newpage
\clearpage
\newpage
\setlength{\unitlength}{1cm}
\begin{figure}
 \includegraphics{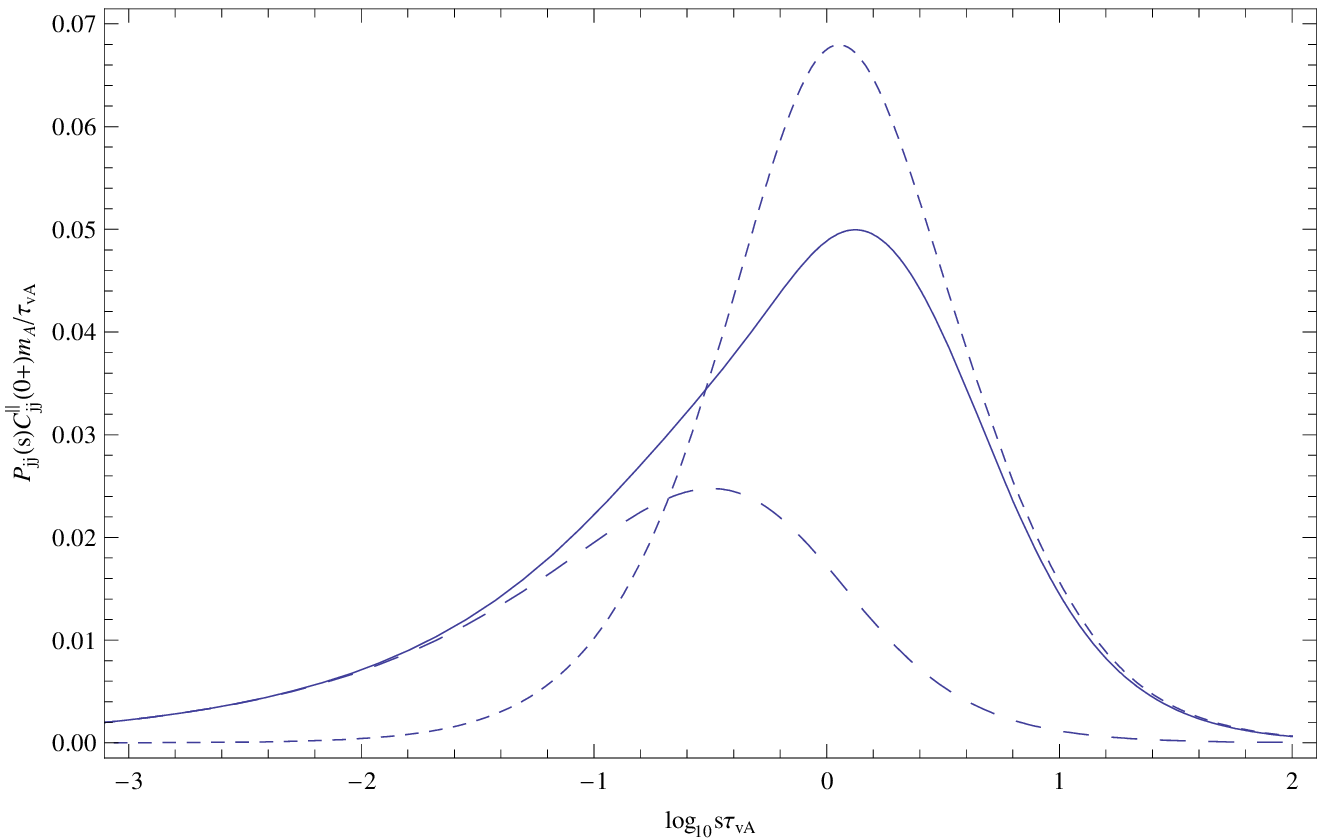}
   \put(-9.1,3.1){}
\put(-1.2,-.2){}
  \caption{}
\end{figure}
\newpage
\clearpage
\newpage
\setlength{\unitlength}{1cm}
\begin{figure}
 \includegraphics{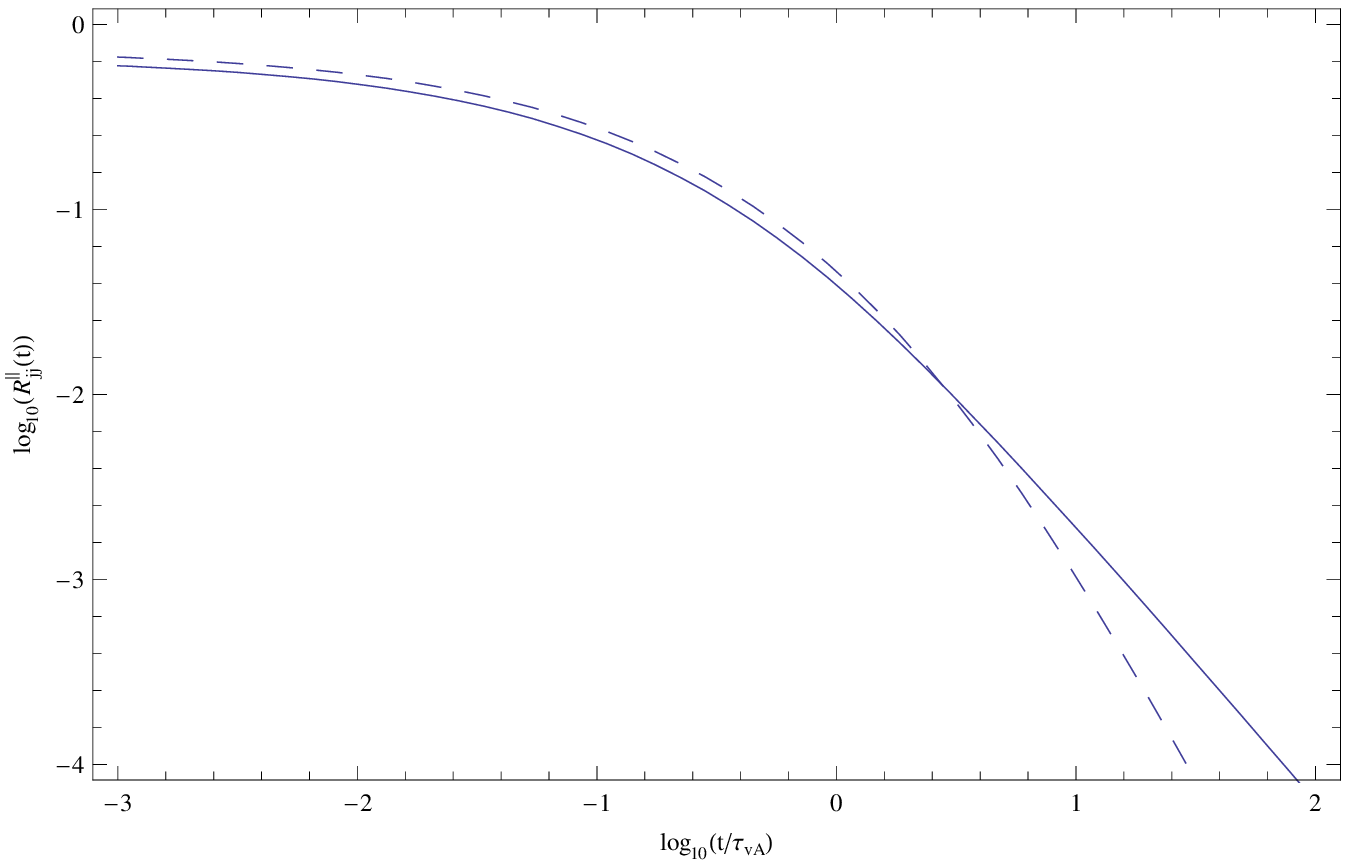}
   \put(-9.1,3.1){}
\put(-1.2,-.2){}
  \caption{}
\end{figure}
\newpage
\end{document}